\documentclass[a4paper,11pt]{article}
\pdfoutput=1 

\usepackage{jcappub} 

\usepackage[T1]{fontenc} 
\usepackage{doi} 
\usepackage{url}
\usepackage{soul}

\usepackage{natbib}
\title{\boldmath Updated constraints on modified gravity from binary pulsars}

\author[a]{S. Bussieres,}
\author[a]{M. Caldarola,}
\author[a,1]{S. Nesseris\note{Corresponding author.}}

\affiliation[a]{Instituto de F\'isica Te\'orica (IFT) UAM-CSIC, Calle Nicol\'as Cabrera 13-15, Campus de Cantoblanco UAM, 28049 Madrid, Spain}

\emailAdd{sofia.bussieres@estudiante.uam.es}
\emailAdd{marienza.caldarola@csic.es}
\emailAdd{savvas.nesseris@csic.es}

\abstract{
Binary pulsars offer a unique natural laboratory to test General Relativity (GR) and probe for deviations from its paradigm, as predicted by alternative theories of gravity. In this paper, we study two such possible deviations: a  time variation of Newton's constant $G$ and the emission of dipolar gravitational radiation.
We use updated data for some well-known pulsars, namely PSR J1738+0333, PSR J1012+5307, and PSR J1713+0747, to extract the Keplerian and post-Keplerian parameters that characterize their orbital dynamics, using recent high-precision pulsar timing data and a Bayesian parameter estimation with Markov chain Monte Carlo (MCMC) techniques. We do this via the \texttt{TEMPO2} software, the  \texttt{MCMC4Tempo2} plugin, and a unified \texttt{python} pipeline to analyze the data.
We then perform a combined analysis of different binary systems to constrain both the time evolution of Newton's constant and the dipolar emission parameter $\kappa_D$. For the best of our three pulsars (PSR J1713+0747), we obtain $\dot{G}/G = (0.32 \pm 0.31) \times 10^{-12}~\text{yr}^{-1}$ at 95\% confidence, along with a stringent constraint on the dipolar emission parameter, $\kappa_D = (-0.04 \pm 0.14) \times 10^{-4}$. Thanks to the recent high-precision timing data sets, we provide updated bounds on key parameters relevant to modified gravity theories, and we find that our results are consistent with GR.}

\begin{document}
\maketitle
\flushbottom
\section{Introduction\label{sec:intro}}
General Relativity (GR), proposed by Albert Einstein in 1915 \cite{Einstein1915}, remains the most successful theory of gravity in modern physics. It describes gravity not as a force, but as the curvature of spacetime caused by mass and energy. GR has passed every experimental and observational test to date, from classic solar system experiments, such as the precession of Mercury’s orbit, gravitational time dilation, and the deflection of light, to modern confirmations through satellite missions like Gravity Probe B \cite{Everitt2011} and the direct detection of gravitational waves from the Laser Interferometer Gravitational-wave Observatory (LIGO) \cite{Abbott2016a}. It has also provided accurate descriptions of extreme environments, such as those near black holes and neutron stars, and underpins our current understanding of the large-scale structure of the universe.

One of the most powerful tools for testing GR in the strong-field and radiative regimes is the study of binary pulsars. These systems consist of at least one neutron star emitting highly regular radio pulses, allowing for precise measurements of orbital parameters over time. The Hulse-Taylor binary pulsar PSR B1913+16 \cite{HulseTaylor1975} was the first to show orbital decay consistent with the emission of gravitational waves, matching GR’s predictions to high accuracy \cite{TaylorWeisberg1982}. Since then, observations of other systems, such as the double pulsar PSR J0737–3039A/B \cite{Kramer_double}, have allowed for even more stringent tests of relativistic effects, including periastron precession, Shapiro delay, and gravitational redshift. Pulsar timing arrays now extend this frontier, aiming to detect nHz-frequency gravitational waves from supermassive black hole binaries and test the consistency of GR on galactic scales.

Despite its empirical success, GR faces significant theoretical challenges. At a fundamental level, it is difficult to reconcile with quantum mechanics, since it does not fit naturally into the framework of quantum field theory. GR is renormalizable only at one loop and non-renormalizable at all loops \cite{Birrell_Davies_1982}, meaning that its predictive power breaks down at very high energies or small distances, such as near the Big Bang or inside black holes. There are also unresolved cosmological issues, such as the cosmological constant problem, why the vacuum energy predicted by quantum theory is so much larger than what is observed \cite{Weinberg1989}, and the mystery of dark energy, which appears to be driving the accelerated expansion of the universe. These challenges suggest that GR may be an effective theory \cite{Birrell_Davies_1982, Weinberg_EFT}, valid up to a point, but requiring modification or extension in a more fundamental framework.

Many alternative theories of gravity have been proposed to address these shortcomings, particularly those that introduce new fields or degrees of freedom. Scalar-tensor theories are among the most well-studied examples \cite{Wagoner1970, Bergmann, BransDicke}. These models extend GR by coupling the metric tensor to one or more scalar fields, often motivated by attempts to embed gravity within a more unified theory like string theory. Scalar fields can modify gravitational dynamics in subtle ways, leading to deviations from GR in both the weak- and strong-field regimes. Notably, in certain scalar-tensor models, neutron stars can undergo ``spontaneous scalarization'', where strong gravitational binding triggers a nonzero scalar field configuration. This can alter the orbital dynamics of binary systems and affect the emission of gravitational waves, making binary pulsars ideal laboratories for testing such effects.

Observational constraints from pulsar timing have already placed tight limits on many modified gravity models. For instance, measurements from the double pulsar system PSR J0737–3039A/B and other relativistic binaries limit the strength and coupling of scalar fields, effectively ruling out large regions of parameter space for some scalar-tensor theories. Similarly, the precise agreement between observed orbital decay rates and GR predictions constrains the emission of dipole gravitational radiation, which would be expected in many alternative models. These constraints are complementary to those obtained from gravitational wave detectors like LIGO and Virgo \cite{Abbott2016, Abbott2019}, which are sensitive to the strong-field, dynamical regime of merging compact objects, but typically over much shorter timescales.

Understanding gravity at both theoretical and observational levels is therefore essential not only for testing Einstein’s theory, but also for probing the nature of spacetime and the fundamental forces. Modifications to GR, whether arising from extra dimensions, string-inspired mechanisms, or new fields, must recover the successes of GR in well-tested regimes while offering explanations for the phenomena it cannot fully address. Binary pulsars provide a rare and valuable opportunity to explore these possibilities with extraordinary precision, helping to constrain or guide the development of more complete theories of gravity.

In our work, we perform Bayesian parameter estimation employing Markov chain Monte Carlo (MCMC) techniques implemented in the \texttt{TEMPO2} software, alongside a unified \texttt{python} analysis pipeline, using high-precision pulsar timing data. This allows us to extract both Keplerian and post-Keplerian parameters that characterize the orbital dynamics of several well-known binary pulsars. We then carry out a joint analysis across multiple systems to constrain the time variation of the gravitational constant ($\dot{G}/G$) and the dipolar gravitational wave emission parameter ($\kappa_D$). The pulsars considered are PSR~J1713$+$0747, PSR~J1738$+$0333 and PSR~J1012$+$5307, with PSR~J0437--4715 as auxiliary binary pulsar, for which, unlike in previous studies, we make no assumptions on the dipolar term, in order to reduce any bias on the time-varying gravitational constant.

The structure of our paper is as follows: Section~\ref{sec:theoretical_framework} contains an overview on some relevant alternative theories beyond GR, including scalar-tensor theories, $f(R)$ models and extra-dimension frameworks. We discuss the corresponding action, and particularly the emergence of a varying gravitational constant through the rescaling of fundamental couplings, but also a possible dipolar gravitational emission, underlining how these effects are related to binary pulsar observables. In Section~\ref{sec:pulsar_timing} we focus on pulsar timing, and how the Time of Arrivals (TOAs) of the pulses are measured and used to determine the Keplerian and post-Keplerian parameters describing the dynamics of a binary system, through a \textit{timing model}. These parameters, particularly the change in orbital period, will be fundamental in constraining deviations from GR.  In Section~\ref{sec:results} we summarize the results obtained through the timing model fitting for our pulsars and the obtained constraints on both the time-varying gravitational constant and the dipolar emission strength. Finally, in Section~\ref{sec:conclusion} we present our conclusions.

\section{Theoretical framework}
\label{sec:theoretical_framework}
To understand how modifications to gravity can be constrained by observations, it is useful to begin with the theoretical structure of GR. In four dimensions, the dynamics of spacetime are governed by the Einstein-Hilbert action:
\begin{equation}
S_{\text{EH}} = \frac{1}{16\pi G_\mathrm{N}} \int d^4x\, \sqrt{-g} \, R + S_{\text{M}},
\end{equation}
where $G_\mathrm{N}$ is Newton's bare gravitational constant, $g$ is the determinant of the spacetime metric $ g_{\mu\nu} $, $ R $ is the Ricci scalar, $ S_{\text{M}} $ corresponds to the action of the matter fields, and we assume $c=1$. In GR, Newton's constant $G_\mathrm{N}$ is a true constant, fixed by experiment, for example Cavendish type experiments. However, in many modified gravity models, especially scalar-tensor theories, $G_\mathrm{N}$ can become effectively dynamical, varying in space or time due to its dependence on additional fields.

Varying the Einstein-Hilbert action with respect to the metric yields Einstein's field equations:
\begin{equation}
R_{\mu\nu} - \frac{1}{2} g_{\mu\nu} R = 8\pi G_\mathrm{N} \,T_{\mu\nu},
\end{equation}
which relate the geometry of spacetime to the energy-momentum content encoded in $ T_{\mu\nu} $. These equations predict a wide range of gravitational phenomena that have been confirmed by observation. Yet, they also assume a specific geometric formulation of gravity and a constant gravitational coupling. Any deviations from these assumptions, such as additional fields, higher dimensions, or modified couplings, lead to new equations of motion and potentially observable differences, especially in strong-field systems like binary pulsars.

Binary pulsars provide one of the most precise astrophysical laboratories for testing gravitational theories. In these systems, a key observable is the change in the orbital period over time, denoted $ \dot{P}_b $. In  GR, the loss of energy due to gravitational wave emission leads to a secular decrease in the orbital period, which for a quasi-circular orbit is given by \cite{Maggiore, PetersMathews1963}:
\begin{equation}
\dot{P}_b = -\frac{192\pi}{5} \left( \frac{2\pi G_\mathrm{N} \mathcal{M} }{P_b} \right)^{5/3} \left(1 + \frac{73}{24}e^2 + \frac{37}{96}e^4 \right) (1 - e^2)^{-7/2},
\label{EinsteinPbdot}
\end{equation}
where $ \mathcal{M} = \frac{(m_1 m_2)^{3/5}}{(m_1 + m_2)^{1/5}} $ is the chirp mass of the system, \( P_b \) is the orbital period, and $ e $ is the Keplerian orbital eccentricity. Precise timing measurements of $ \dot{P}_b $ allow for stringent tests of GR and provide constraints on alternative theories that predict different rates of energy loss, such as those involving scalar radiation or modifications to the propagation of gravitational waves.

\subsection{Modified gravity}
Despite the remarkable success of GR, the problems discussed in the first section raise questions about whether it truly represents the fundamental theory of gravity.
This questioning is not new. In fact, just after the publication of GR, physicists already began to explore possible extensions  to incorporate it in a larger and more unified theory. 

Some of the first notable proposals were carried by Eddington \cite{Eddington:1924} and Weyl  \cite{Weyl:1918}. Eddington's theory of connections would later inspire metric-affine and Palatini formulations of gravity \cite{OLMO_2011}, and his attempts to derive physical constants from geometric principles would also influence Dirac, who conjectured that Newton's constant might vary with time \cite{Dirac:1937}. This idea was explored years later by Brans and Dicke in the context of scalar-tensor theories of gravity \cite{BransDicke}. Weyl attempted to unify gravitation and electromagnetism through a theory based on local scale invariance. Although inconsistent, Weyl’s model introduced the idea of gauge symmetry, which would later become foundational in modern field theory. 

These early attempts already hinted at the possibility that GR might be just one piece of a broader theoretical framework. Lovelock's theorem \cite{Lovelock:1972vz} establishes that in four dimensions the only rank-2 tensor that can be derived from an action depending on the metric and its first two derivatives, and that leads to second-order field equations, is the Einstein tensor (plus a cosmological constant term). This result highlights the rigidity of GR and suggests that any extension of the theory must go beyond at least one of these assumptions, that is, by adding new fields, higher curvature terms or extra dimensions. 

\subsubsection{Scalar-Tensor theories}
Scalar-tensor theories represent one of the most common ways to study deviations from GR. The relatively simple structure of their field equations allows one to find analytical solutions more easily than in other theories. 
In the Jordan frame, the scalar-tensor action is given by \cite{Bergmann, clifton2012modified}:
\begin{equation}
    S = \frac{1}{16\pi G_\mathrm{N}} \int d^4x\,\sqrt{-g} \left[ \phi R - \frac{\omega(\phi)}{\phi} \nabla_\mu \phi \nabla^\mu \phi - 2\Lambda(\phi) \right] + S_\mathrm{M}(\psi, g_{\mu\nu}),
    \label{lagrascalartensor}
\end{equation}
where $\omega$ and $\Lambda$ are functions of the scalar field $\phi$. This action defines an effective gravitational constant, denoted as $G_{\text{eff}}$, that scales approximately as \begin{equation}
    G_{\text{eff}} \, \propto \phi^{-1},
    \label{gdotSCALARTENSOR}
\end{equation}
that is, it is no longer a constant but depends on the dynamical degree of freedom introduced by the scalar field.

\subsubsection{$f(R)$ theories}
An alternative way to deviate from GR is by allowing the field equations to contain derivatives of higher than second order. 
Such theories often exhibit rich phenomenology and may improve the renormalizability properties of gravity.
Working with higher-order derivatives typically raises concerns about instabilities, particularly the emergence of ghost-like degrees of freedom. However, some specific theories, such as $f(R)$ gravity, are less susceptible to them. This is because the higher-order derivatives act on modes that are non-dynamical in GR, effectively promoting them to dynamical fields rather than introducing ghosts. 

The $f(R)$ theories are fourth-order theories of gravity and represent one of the simplest extensions of GR. They are constructed by modifying the gravitational action by promoting the Ricci scalar $ R $ in the Einstein-Hilbert Lagrangian to a general (usually analytic) function  $ f(R)$. These theories have been extensively studied as toy models to explore deviations from GR \cite{Sotiriou_2010, De_Felice_2010, Nojiri_2011}, especially after the realization that certain models can drive a phase of early accelerated expansion of the Universe \cite{STAROBINSKY198099}. In this framework, the Einstein-Hilbert action is replaced by the general action
\begin{equation}
S=\frac{1}{16 \pi G_\mathrm{N}}\int{d^4 x\,\sqrt{-g} \,f(R)}+S_\mathrm{M}(g_{\mu \nu}, \psi),
\label{modaction}
\end{equation}
The field equations can be written in terms of the Einstein tensor as 
\begin{equation}
G_{\mu \nu}
= \frac{8 \pi G_\mathrm{N} }{f_R} T_{\mu \nu} +  \frac{f- R f_R}{2 f_R}g_{\mu \nu}
+ \frac{\nabla_{\mu} \nabla_{\nu} f_R - g_{\mu \nu} \Box f_R}{f_R}.
\label{frfield}
\end{equation}
where $\partial f(R)/ \partial R \equiv f_R $.
The trace of Eq. \eqref{frfield} manifests the dynamical nature of the new degree of freedom, which can also be seen through the equivalence of the metric $f(R)$ gravity to $\omega=0$ Brans-Dicke theory with $\Lambda \rightarrow 0$ in Eq.~\eqref{lagrascalartensor}. That is, it effectively reduces to a case of scalar-tensor theories and again, the effective gravitational constant depends on the additional dynamical degree of freedom as 
\begin{equation}
    G_{\text{eff}} \, \propto \, f_R^{-1},
\end{equation}
under the condition that $f_R >0$ to ensure the positivity of $G_{\text{eff}}$.

\subsubsection{Extra dimensional theories}
It appears to be a generic feature of higher-dimensional gravity theories, and of string-inspired models in particular, that their low-energy effective descriptions in four dimensions involve dynamical couplings. That is, quantities such as the gravitational constant or gauge couplings are no longer fixed, but become dependent on scalar fields arising from the extra-dimensional geometry \cite{UZAN}. Considering a simple case of one extra dimension as in Kaluza Klein models, the five-dimensional metric can be decomposed in 4D fields as
\begin{equation}
\bar{g}_{AB} = \begin{pmatrix}
g_{\mu\nu} + \dfrac{A_\mu A_\nu}{M^2} \phi^2 & ~~\dfrac{A_\mu}{M} \phi^2 \\
\dfrac{A_\nu}{M} \phi^2 & \phi^2
\end{pmatrix},
\label{gAB}
\end{equation}
where $g_{\mu\nu}$ is the four-dimensional metric, $A_\mu$ is the vector field identified with the electromagnetic potential, $\phi$ is a scalar field associated with the size of the compactified extra dimension, and $M$ is a mass scale. Then, the five-dimensional Einstein-Hilbert action reads \cite{UZAN}
\begin{equation}
    S = \frac{1}{12\pi^2 G_5} \int \mathrm{d}^5x\,\sqrt{-\bar{g}} \,\bar{R},
\end{equation}
where $G_5$ is the gravitational constant in five dimensions. Upon dimensional reduction and integration over the compact fifth dimension ($x^4 \equiv y$), the effective four-dimensional action becomes
\begin{equation}
    S = \frac{1}{16\pi G_\mathrm{N}} \int\,\mathrm{d}^4x\,\sqrt{-g}\, \left( R - \frac{\phi^2}{4M^2} F_{\mu\nu} F^{\mu\nu} \right) \phi ,
    \label{GDOTEXTRADIM}
\end{equation}
where $G_\mathrm{N}=3\pi G_5/4 \int dy$ defines the effective 4D gravitational constant and $F_{\mu\nu}$ is the field strength tensor of the vector field. 

The scalar field $\phi$ introduced by the extra dimension couples non-minimally to both the gravitational sector (assuming a Jordan frame) and the electromagnetic sector, through the gauge kinetic term.\footnote{This coupling to $F_{\mu\nu} F^{\mu\nu}$ cannot be eliminated by a conformal transformation due to the conformal invariance of electromagnetism in 4D. See Ref.~\cite{UZAN} for a detailed discussion.} As a result, $\phi$ controls the effective values of fundamental couplings. In particular, both the gravitational constant and the gauge coupling become dynamical fields that scale as
\begin{equation}
     G_\mathrm{eff} \propto \phi^{-1}, \qquad g_{\text{YM}}^{-2} \propto \frac{\phi^2}{G} \propto \phi^3.
\end{equation}
This result generalizes to compactifications involving $D$ extra spatial dimensions, yielding a gravitational constant that scales as $G_\mathrm{eff} \propto \phi^{-D}$~\cite{CREMMER197761}.

In what follows, in order to simplify the notation we will refer to the potentially evolving Newton's constant $G_\mathrm{eff}$ as just $G$, stressing however, that this is different from the bare laboratory value of $G_\mathrm{N}$.

\subsection{Tests of GR with pulsars}
It has been pointed out that a time-varying gravitational constant would affect the dynamics of binary pulsars, producing  periodic modulations in the observed pulse period, as studied within the Brans-Dicke framework by Ref. \cite{Eardly}. Later, Ref. \cite{Nordtvedt} stated that a changing $G$ would alter the compactness and mass of the pulsar, introducing an additional contribution to the derivative of the orbital period. For a weakly self-gravitating companion, this contribution can be written, at leading order, as
\begin{equation}
\dot{P}_b^{\dot{G}} \simeq -2 \frac{\dot{G}}{G} \left[ 1 - \frac{2M_p + 3M_c}{2(M_p + M_c)} s_p - \frac{2M_c + 3M_p}{2(M_p + M_c)} s_c \right] P_b,
\label{Nordvelt}
\end{equation}
where \( M_p \) and \( M_c \) are the masses of the pulsar and companion, respectively, and \( s_p \), \( s_c \) denote their sensitivities to changes in the gravitational constant \cite{Will_1993}
\begin{equation}
    s_p \equiv -\left. \frac{\partial \ln M_p}{\partial \ln G} \right|_N 
\quad \text{and} \quad 
s_c \equiv -\left. \frac{\partial \ln M_c}{\partial \ln G} \right|_N,
\label{sensitivities}
\end{equation}
where $N$ is the number of baryons and it is held fixed. These values depend on the theory of gravity, the form of the equation of state and the mass of the pulsar. 

Due to the complicated internal composition of neutron stars, various equations of state (EoS) have been investigated in the literature using a semi-analytical, semi-numerical approach. Following the conclusions of Ref.~\cite{EspositoDamour_1992}, we adopt, as a first-order approximation, that the sensitivity $s_p$ scales linearly with the stellar mass. In particular, for the AP4 equation of state in Ref.~\cite{Lattimer2001}, a typical value of the sensitivity for a $1.33\,M_\odot$ neutron star is $s_p \simeq 0.16$~\cite{Zhu_2018}. Hence, we take
\begin{equation}
    s_p = 0.16 \left(\frac{M_p}{1.33\,M_\odot}\right).
\end{equation}

For weakly self-gravitating bodies, such as dwarf companions, the sensitivity can be approximated as $s_c \simeq -E_{c, \text{grav}} / M_c \,c^2$, where \( E_{c, \text{grav}} \) is the gravitational binding energy. Typical white dwarfs have masses in the range 
$ 0.5-0.7 \, M_\odot $ \cite{keplerWD} and radii between $ 0.008-0.02 \, R_\odot $ \cite{ShipmanWD}, yielding sensitivity values on the order of \( s_c \sim 10^{-5} \) to \( 10^{-4} \), which are typically negligible compared to neutron star sensitivities.
This way, for a binary system formed by a pulsar and a white dwarf companion, we approximate Eq.~\eqref{Nordvelt} to be
\begin{equation}
\dot{P}_b^{\dot{G}} \simeq -2 \frac{\dot{G}}{G} \left[ 1 - \left(1+\frac{3M_c}{2M}\right) s_p \right] P_b,
\label{Nordveltsimp}
\end{equation}
where $M$ is the total mass of the system. 

On the other hand, it has been previously discussed that in gravitational theories that introduce scalar degrees of freedom, such as scalar-tensor theories, the gravitational interaction is mediated not only by the metric tensor but also by one or more scalar fields. This leads to violations of the \textit{strong equivalence principle} (SEP) and also introduces new terms in the post-Newtonian (PN) expansion. In particular, one finds the emergence of dipolar gravitational radiation at 1.5PN order~\cite{Eardly}, while in GR the first allowed radiation term appeared only at 2.5PN order through quadrupole emission.
In the binary system, this dipolar component is proportional to the square of the difference in gravitational sensitivities of the two bodies, with sensitivities defined as, if $\phi$ is the additional degree of freedom,
\[
    s_p \equiv \left. \frac{\partial \ln M_p}{\partial \ln \phi} \right|_N 
\quad \text{and} \quad 
s_c \equiv -\left. \frac{\partial \ln M_c}{\partial \ln \phi} \right|_N,
\]
which is equivalent to Eq.~\eqref{sensitivities} remembering that in scalar-tensor theories the effective gravitational constant scales as \( G \propto \phi^{-1} \).
If the binary system consists of bodies with significantly different sensitivities, such as a pulsar–white dwarf binary, this effect becomes more relevant. Moreover, since it arises at 1.5PN order, it corresponds to corrections of order \( (v/c)^3 \) in the equations of motion, and can thus dominate the energy loss in the system. Since its impact is particularly significant in systems with strongly asymmetric sensitivities~\cite{Mirshekari}, pulsar-dwarfs systems represent excellent laboratories for testing scalar-tensor theories.

Since the presence of dipolar radiation contributes to the energy loss of the binary system, it will also introduce an additional decay in the orbital period. For white dwarf companions, at leading order \cite{DamourFarese_1996, Will_1993},
\begin{equation}
\dot{P}_b^{\text{D}} \simeq -4\pi^2 \, \frac{T_\odot M_c}{P_b^2} \, \frac{q}{q + 1} \, \kappa_D\,s_p^2,
\label{dipolar}
\end{equation}
where \( T_\odot = G_\mathrm{N} M_\odot/c^3 = 4.9255\,\mu\text{s} \), the mass rate is $q=M_p/M_c$ and \( \kappa_D \) is a dimensionless constant that depends on the specific gravity theory being considered but is independent of the internal structure of the bodies \cite{Will_1993}.

Both these potential time-variation of the gravitational constant and the presence of dipolar radiation stand out among the effects that deviate from GR, particularly within scalar-tensor theories. In this work, we aim to constrain these effects, and consequently the theories that predict them, through the study of binary pulsars and their observables.

\section{Pulsar timing and methodology}
\label{sec:pulsar_timing}
Radio pulsars were identified with rotating neutron stars shortly after the first pulsar PSR B1919+21 was discovered in 1967 \cite{HEWISH1968, Gold1968}. The strong magnetic field characterising these compact objects produces a radiation cone along the axis of the field, which is usually not aligned with the rotating axis \cite{lorimer2005handbook}. This beam sweeps across our line of sight, making possible its detection from Earth through radio telescopes.
The rotational stability of these objects ensures a remarkable precision of the emission of the pulses, allowing to detect any deviation in the expected arrival time of the pulses, or  TOAs. This technique, known as \textit{pulsar timing}, has turned pulsars into exceptional laboratories for studying the evolution of astrophysical systems and for conducting precise tests of gravity, as well as research in plasma physics and nuclear physics, among other fields \cite{NSplasmanuc}.

Pulsar observations are typically conducted at radio frequencies between a few hundred MHz to a few GHz. Due to their intrinsically weak emission, individual pulses are frequently indistinguishable from the background noise. To overcome this, the time series of pulses is commonly integrated or `folded' over both time and frequency to increase the signal-to-noise ratio (SNR), averaging out Gaussian noise and revealing a stable integrated pulse profile. The shape of individual pulses can vary significantly, so integrating hundreds or thousands of pulses is necessary to obtain a reproducible and high-SNR profile \cite{Hu_2024}.

To measure accurate TOAs, the integrated profile is cross-correlated with a standard or template profile at the same observing frequency. 
However, the pulse period may not be very easily predicted from the pulse TOA intervals, especially if the pulsar happens to be in a binary system. The propagation effects on signals and the motion of pulsars, especially if they are orbiting with a companion, cause various time delays between the pulse emitted and the TOAs. The goal of pulsar timing is to develop a model of the pulse phase as a function of time, so that all future pulse arrival times can be predicted with a good degree of accuracy.

\subsection{Timing model fitting}
The pulses of a particular pulsar are measured over many years at radio observatories, so they are initially referenced to local clocks. 
In order to interpret them in a meaningful astrophysical context, it is necessary to transform them into an inertial reference frame, that is, reconstruct the pulse arrival times in the pulsar co-moving frame.
This is done firstly by referencing the measured pulses 
in the Solar System barycenter (SSB) frame. These corrections account for a wide range of effects, including kinematic contributions due to Earth's motion or relativistic phenomena such as gravitational time delays.
Once the TOAs are referenced to the SSB, further adjustments may be necessary if the pulsar is part of a binary system. In this case stronger gravitational effects need to be included in the dynamics of the binary system.

To do so, a \textit{timing model} (or \textit{timing ephemeris}) is fitted to the TOAs. This model describes the pulsar's rotational behavior, including its spin frequency and spin-down rate, as well as astrometric parameters like position, proper motion, and parallax. In binary systems, the timing model also includes parameters characterizing the orbital motion, such as secular changes due to GW emission. The fitting process seeks to minimize the timing residuals, defined as the differences between the observed TOAs and those predicted by the model. Any systematic trends in these residuals may indicate deficiencies in the model, such as unmodeled effects or a poor estimation of the astrometric or rotational parameters. In exceptional cases, they may serve as detectors of new physical phenomena, making pulsar timing an extraordinarily sensitive tool for tests of fundamental physics. 

The intrinsic rotation frequency of a pulsar can be expanded as 
\begin{equation}
\nu(T) = \nu_0 + \dot{\nu}_0 (T - t_0) + \frac{1}{2} \ddot{\nu}_0 (T - t_0)^2 + \ldots,
\label{spinrot}
\end{equation}
where $T$ is the pulsar proper time and $\nu_0\equiv \nu(t_0)$ is the spin frequency at the reference epoch $t_0$. The radiation of a pulsar carries away its rotational energy, leading to a gradual decrease in frequency described by $\dot{\nu}_0$ and higher-order derivatives such as $\ddot{\nu}_0$.

The accumulated phase of the pulsar, denoted by $\Phi(T)$, tracks the number of full rotations since a reference epoch. It can be obtained by integrating the frequency over time 
\begin{equation}
\frac{1}{2 \pi}\Phi(T) = \int_0^T \nu(T')\, dT' = \nu_0 T + \frac{1}{2} \dot{\nu}_0 T^2 + \frac{1}{6} \ddot{\nu}_0 T^3 + \cdots.
\label{phase_acc}
\end{equation}
This accumulated phase determines the number of pulse cycles since emission occurs when the phase satisfies $\Phi(T_n) = \Phi_0 + 2\pi n$ for some integer $n$. Therefore, the proper times between successive pulse emissions can be inferred by inverting the phase function.

As can be seen,  Eq.~\eqref{spinrot} is already referring to the pulsar comoving frame. However, since the pulses we measure are in the non-inertial frame of the radio telescopes, time delays must be added to the measured TOAs when translating them to the SSB \cite{DDtiming, TaylorWeisberg1989}. Following the notation and expressions in \cite{Maggiore}
\begin{equation}
t_{\mathrm{SSB}} = \tau_{\mathrm{obs}} + \Delta_C - \frac{D}{\nu^2} + \Delta_{E\odot} + \Delta_{R,\odot} + \Delta_{S,\odot}.
\label{timingssb}
\end{equation}
For an isolated pulsar, the proper time of the pulsar ($T$) corresponds to the arrival time at the SSB ($t_\mathrm{SSB}$) of infinite observing frequency, up to an additive constant.\footnote{This constant represents an unknown offset due to the imprecise knowledge of the distance between the pulsar and the SSB. It is not relevant in practice, since pulsar timing relies only on differences between TOAs \cite{Maggiore}.} Among the different time corrections considered in Eq.~\eqref{timingssb} we find the corresponding clock corrections ($\Delta_C$) that unify measurements by different observatories in a common reference timescale, the Roemer delay ($\Delta_{R, \odot}$) due the geometrical position of the observatory-Earth-Sun system, the Shapiro delay ($\Delta_{S, \odot}$) caused by the gravitational potential of the Sun, the Einstein delay ($\Delta_{E, \odot}$) relating proper and coordinate times of emission, and a time delay due the dispersion of the signal in the interstellar medium ($D$).  

Once the TOAs are translated to the SSB, additional corrections ($\Delta_B$) need to be considered if the pulsar is present in a binary system. These describe the orbital motion of the pulsar and the gravity field of the companion, and translate the measured time to the pulsar's proper time, where pulses are assumed to be emitted at regular intervals.
\begin{equation}
    t_\mathrm{SSB} = T + \Delta_B.
\end{equation}
This relationship is inverted during model fitting: the parameters of the binary contributing in $\Delta_B$ are adjusted so that the predicted emission times match the observed TOAs. Through the phase model of the pulsar's spin evolution given in Eq.~\eqref{spinrot}, which tracks the accumulated number of rotations since a reference epoch, we can reconstruct the time intervals between successive emissions in the pulsar’s own proper time, even if we do not know the absolute value $T$ of the pulsar at given moment.

Finally, the complete timing formula relating the proper time of the pulsar to the measured in the laboratory clocks, departing from Eq.~\eqref{timingssb}, reads \cite{DDtiming}
\begin{equation}
T = \tau_{\mathrm{obs}} + \Delta_C - \frac{D}{\nu^2} + \Delta_{E\odot} + \Delta_{R,\odot} + \Delta_{S,\odot} - \left( \Delta_{EB} + \Delta_{RB} + \Delta_{SB} + \Delta_A\right).
\label{timingCOMPLETE}
\end{equation}

A similar approach to that used for the Earth–Sun system has been applied to compute the time corrections related to the pulsar-companion system $\Delta_B= \Delta_{EB} + \Delta_{RB} + \Delta_{SB} + \Delta_A$. Likewise, we account for a Roemer, Shapiro and Einstein delays for the binary system.
The  main difference now is that binary systems involve much stronger gravitational fields. While this complicates severely the modeling, it also allow us to measure the relativistic effects in such systems, as GWs emission.
Moreover, an aberration delay is included ($\Delta_A$) placing corrections due to the periodic changes in the direction of pulse emission while the pulsar follows its binary motion. These corrections depend on three types of parameters.

First, the parameters that describe the pulsar itself, such as its proper motion, the initial phase $\phi_0$, right ascension $\alpha$ and declination $\delta$, as well as the spin frequency $\nu$ and its derivative $\dot{\nu}_0$, which characterizes the spin-down of the pulsar. Additionally, the orbital motion in binary pulsar systems can be described by five \textit{Keplerian} parameters,
\begin{equation}
    \{P_b, T_0, x_0, e_0, \omega_0\},
\end{equation}
where $P_b$ is the orbital period, and $T_0$ is the time of periastron passage used as a reference epoch. The quantity $x_0$ is the projected semi-major axis of the pulsar's orbit, $e_0$ is the orbital eccentricity, and $\omega_0$ is the longitude of periastron. All these parameters are evaluated at $T = T_0$.

So far, the parameters described above correspond to a purely Newtonian parametrization of the orbital motion.
In order to properly account for the time delay terms associated with relativistic binary effects, as included in Eq.~(\ref{timingCOMPLETE}), a set of independently measurable \textit{post-Keplerian} (PK) parameters must be introduced
\begin{equation}
    \{\dot{\omega}, \gamma, \dot{P}_b, r, s, \delta_\theta, \dot{e}, \dot{x}\},
    \label{ppk}
\end{equation}
where $\dot{\omega}$ is the advance rate of periastron, $\dot{P}_b$ is the decay rate of the orbital period, $\gamma$ is the amplitude of Einstein delay, $r$ and $s$ are called the range and shape of Shapiro delay, $\delta_\theta$ is a relativistic deformation of the orbit and $\dot{x}$ and $\dot{e}$ describe the secular changes in the projected major axis and eccentricity.

The parameters in Eq.~\eqref{ppk} are the most commonly used within the so-called parametrized post-Keplerian (PPK) formalism, a phenomenological parametrization proposed by \cite{DDcelestialmec, DDcelestialmec2}, and later extended by \cite{DDtiming}. These quantities can be extracted from the pulsar timing and are not dependent on the predictions of any specific Lorentz-invariant gravity theory at the 1PN order \cite{Hu_2024}. Other non-separately measurable PK parameters are 
\begin{equation}
    \{\delta_r, A, B\},
\end{equation}
where $\delta_r$ is also a parameter describing a relativistic deformation of the orbit, and $A$ and $B$ are aberration parameters. Considering GWs radiation, a Kepler-like equation was proposed by \cite{DDcelestialmec, DDcelestialmec2} as 
\begin{equation}
    u - e \sin u = 2\pi \left[ \left( \frac{T - T_0}{P_b} \right) - \frac{\dot{P}_b}{2} \left( \frac{T - T_0}{P_b} \right)^2 \right],
\end{equation}
where $u$ is the eccentric anomaly, related to the true anomaly $A_e$ and the longitude of periastron $\omega$ via
\begin{equation}
    A_e(u) = 2 \arctan \left[ \left( \frac{1+e}{1-e} \right)^{1/2} \tan \frac{u}{2} \right],
\end{equation}
\begin{equation}
    \omega = \omega_0 + \left( \frac{P_b \dot{\omega}}{2\pi} \right) A_e(u).
\end{equation}
We are now in position to derive the expressions for the effects included in the timing formula Eq. (\ref{timingCOMPLETE}). Respectively, the Roemer, Einstein, Shapiro \cite{ShapiroB} and the aberration \cite{lorimer2005handbook} delays for a binary are given by

\begin{equation}
    \Delta_{RB} = x \sin\omega (\cos u - e(1 + \delta_r)) 
+ x (1 - e^2 (1 + \delta_\theta)^2)^{1/2} \cos\omega \sin u,
\end{equation}
\begin{equation}
    \Delta_{EB} = \gamma \sin u,
\end{equation}
\begin{equation}
    \Delta_{SB} = -2r \ln \left\{ 1 - e \cos u - s \left[ 
\sin\omega (\cos u - e) + (1 - e^2)^{1/2} \cos\omega \sin u 
\right] \right\},
\end{equation}
\begin{equation}
\Delta_{\mathrm{A}} = A \left\{ \sin\left[\omega + A_e(u)\right] + e \sin \omega \right\} + B \left\{ \cos\left[\omega + A_e(u)\right] + e \cos \omega \right\}.
\end{equation}
Once we have derived each term present in the timing formula, the orbital parameters are obtained through the model fitting by pulsar timing in a completely phenomenological manner, that is, we do not assume any particular theory of gravity.
In GR, these PPK parameters are functions of the Keplerian parameters and the masses of the system, such as $\dot{P_b}$ in  Eq.~\eqref{EinsteinPbdot}. This implies that knowing the value of the Keplerian parameters and the masses, the PK parameters extracted from the fit can provide a test of GR. 

\subsection{Corrections to change in orbital period}
Regarding our analysis, we extract a preliminary estimate of the orbital period variation, $\dot{P}_b^{\text{obs}}$, based on the timing model. This quantity is inferred from the observational data and the parameters describing the pulsar's position.
This parameter is affected not only by the loss of energy due to GW emission, but also by other external effects, primarily of kinematic origin.
The quantity $\dot{P}_b^{\text{obs}}$ is therefore an apparent orbital period change. 

In order to compare it to the theoretical relativistic prediction derived in Eq. \eqref{EinsteinPbdot}, we must take into account all the different contributions that shift it away from the intrinsic value, $\dot{P}_b^{\text{int}}$ \cite{DamourTaylor1991}.
\begin{equation}
\left( \frac{\dot{P}_b}{P_b} \right)^{\text{obs}} = \left( \frac{\dot{P}_b}{P_b} \right)^{\text{int}} + \left( \frac{\dot{P}_b}{P_b} \right)^{\text{kin}} +\left( \frac{\dot{P}_b}{P_b} \right)^{\text{accel}}+\left( \frac{\dot{P}_b}{P_b} \right)^{\dot{M}_p}+
\left( \frac{\dot{P}_b}{P_b} \right)^{\dot{M}_c}+ \cdots
\label{Obs-Int}
\end{equation}
The most relevant and historically effects discussed in the literature are primarily due to mass loss from the pulsar or from its companion ($\dot{P}_b^{\dot{M}_c}, \, \dot{P}_b^{\dot{M}_p}$), fluctuations in the Galactic acceleration caused by local clustering of accelerating centers near the Sun or the pulsar ($\dot{P}_b^{\text{accel}}$), and ephemeris uncertainties related to the Earth's position with respect to the SSB, among others. However, as discussed in Ref. \cite{DamourTaylor1991}, all these effects are negligible compared to the kinematic effects of Galactic origin, and can therefore be safely approximated as 
\begin{equation}
\left( \frac{\dot{P}_b}{P_b} \right)^{\text{obs}} \simeq  \left( \frac{\dot{P}_b}{P_b} \right)^{\text{int}} + \left( \frac{\dot{P}_b}{P_b} \right)^{\text{kin}}.
\end{equation}
This correction arises from the Doppler shift caused by the relative motion between the pulsar and the Sun, which slightly alters the intrinsic orbital period according to
\begin{equation}
P_b^{\text{obs}} = P_b^{\text{int}} \left( 1 + \frac{\mathbf{v} \cdot \mathbf{n}}{c} \right),
\label{Pdoppler}
\end{equation}
where $\mathbf{v}$ is the pulsar's velocity relative to the Solar System, and $\mathbf{n}$ is the unit vector along the line of sight. Although this deviation is small enough to be physically negligible (i.e., we may safely work with $P_b^{\text{obs}} \equiv P_b$), the associated Doppler-induced accelerations are significant. Differentiating Eq. \eqref{Pdoppler} and separating the line-of-sight and transverse contributions yields
\begin{equation}
\dot{P}_b^{\text{obs}} = \dot{P}_b^{\text{int}} + \left( \frac{\mathbf{a} \cdot \mathbf{n}}{c} + \frac{\mu^2 d}{c} \right) P_b,
\label{Pdotdoppler}
\end{equation}
where the vector $\mathbf{a}$ represents the relative acceleration of the pulsar with respect to the Solar System, $\mathbf{n}$ is the unit vector pointing along the line of sight, $d$ is the pulsar’s distance, and $\mu$ is the total proper motion. 
 
The first term in the parentheses in Eq. \eqref{Pdotdoppler} arises from the differential acceleration between the pulsar system and the Solar System, projected along the line of sight to the pulsar. Following the expressions originally given by Refs. \cite{NiceTaylor1995b, DamourTaylor1991}, which include only the Galactic acceleration in the plane of the Galaxy, we extend the model to also include the vertical component of the Galactic potential, as discussed in Ref. \cite{Lazaridis}. The full Galactic acceleration contribution can therefore be written as
\begin{equation}
\dot{P}_b^{\text{gal}} \equiv \frac{\mathbf{a} \cdot \mathbf{n}}{c} P_b = 
- \frac{K_z |\sin b|}{c} P_b 
- \frac{\Omega_{\odot}^2 R_{\odot}}{c} \cos b \left( \cos l + \frac{\beta}{\beta^2 + \sin^2 l} \right) P_b,
\label{pbgalact}
\end{equation}
where $\Omega_{\odot} = 27.2 \pm 0.9$ km\,s$^{-1}$\,kpc$^{-1}$ \cite{feastwhitelock} is the Galactic angular velocity at the Sun's position, $R_{\odot} = 8.0 \pm 0.4$ kpc \cite{eisenhauer} is the Galactocentric distance of the Sun, $b$ and $l$ are the Galactic latitude and longitude of the pulsar, and $
\beta \equiv (d/R_0) \cos b - \cos l$. 

The first term on the right-hand side of Eq.~\eqref{pbgalact} accounts for the vertical acceleration of the pulsar with respect to the Galactic plane. To compute this term, we follow the prescription of Ref.~\cite{HolmbergFlynn2004}, who provides an empirical approximation for the vertical component of the Galactic gravitational field, valid for heights \( z_{\text{kpc}} \equiv |d \sin b| \leq 1.5 \, \text{kpc} \) as
\begin{equation}
K_z (10^{-9}~\text{cm}~\text{s}^{-2}) \simeq 2.27\, z_{\text{kpc}} + 3.68 \left(1 - e^{-4.31\, z_{\text{kpc}}} \right).
\end{equation}

The second term in Eq.~\eqref{pbgalact} represents the contribution from differential Galactic rotation within a flat rotation curve model \cite{DamourTaylor1991}, while the last term in Eq.~\eqref{Pdotdoppler} represents the so-called Shklovskii effect \cite{Shklovskii1970}, which arises from the apparent acceleration caused by the pulsar’s transverse motion with respect to the observer. This kinematic effect is purely geometric in nature and is equivalent to a centrifugal acceleration associated with the pulsar's proper motion. It can be expressed as
\begin{equation}
\dot{P}_b^{\mathrm{Shk}} \equiv \left( \mu_\alpha^2 + \mu_\delta^2 \right) \frac{d}{c} P_b,
\label{shkolvs}
\end{equation}
where \( \mu_\alpha \) and \( \mu_\delta \) are the proper motion components in right ascension and declination (in angular units, typically mas\,yr\(^{-1}\)). The total proper motion is \( \mu = \sqrt{\mu_\alpha^2 + \mu_\delta^2} \), and thus \( \mu^2 d / c \) represents the apparent line-of-sight acceleration due to the pulsar's transverse velocity.

Finally, once we obtained the intrinsic value $\dot{P}_b^{\mathrm{int}}=\dot{P}_b^{\mathrm{obs}}-\dot{P}_b^{\mathrm{Shk}}-\dot{P}_b^{\mathrm{Gal}}$, we can compare it to the predicted value by GR, $\dot{P}_b^{\mathrm{GR}}$, given by Eq. \eqref{EinsteinPbdot}. The excess between these two quantities $\dot{P}_b^{\mathrm{exc}}=\dot{P}_b^{\mathrm{int}}-\dot{P}_b^{\mathrm{GR}}$ allows us to test phenomena beyond GR.
Since the alternative theories of gravity we have discussed previously predicted a varying gravitational constant and dipolar emission, both of these effects could be responsible of any deviation from the GR prediction, $\dot{P}_b^{\mathrm{exc}}=\dot{P}_b^{\mathrm{D}}+\dot{P}_b^{\dot{G}}$ with the expressions discussed previously in Eqs.\eqref{dipolar} and \eqref{Nordveltsimp}, leading to
\begin{equation}
    \frac{\dot{P}_b^\mathrm{exc}}{ P_b} = -2 \frac{\dot{G}}{G} \left[ 1 - \left(1+\frac{3M_c}{2M}\right) s_p \right]-4\pi^2 \, \frac{T_\odot M_c}{P_b^2} \, \frac{q}{q + 1} \, \kappa_D \,s_p^2.
    \label{sistemadeeq}
\end{equation}

\noindent This will be the central relation we will use to constrain both the effects of a varying gravitational constant and dipolar emission via binary pulsar analysis.

\subsection{Numerical pipeline}
Both the conversion from the measured TOAs to barycentric arrival times and the model fitting required to obtain precise pulsar parameters are complex and can only be carried out computationally. The most widely used package performing this process is \texttt{TEMPO2}, a new version of the original \texttt{TEMPO} software distributed by Princeton University and the Australia Telescope National Facility \cite{Hobbs_TEMPO2, EDWARDS}. This version incorporates upgrades such as improved models, higher precision and support for relativistic binary effects.

\texttt{TEMPO2} analyses the TOAs from observations through the timing model, a solar system ephemeris and clock information from the observatories. 
The software then fits for the parameters of the timing model, using a weighted least-squares algorithm, minimizing 
\begin{equation}
    \chi^2 = \sum_{i=1}^{N} \left( \frac{R_i}{\sigma_i} \right)^2,
\end{equation}
where $N$ is the number of observations, $\sigma_i$ is the uncertainty of the $i$th TOA and $R_i$ is the pre-fit residual given by
\begin{equation}
    R_i = \frac{\phi_i - N_i}{\nu},
\end{equation}
where $\phi_i$ is the predicted accumulated phase based on the model pulse frequency, see Eq.~\eqref{phase_acc}, and $N_i$ is its nearest integer, representing the expected number of whole pulsar rotations. These residuals reflect the phase mismatch between the predicted and observed pulse, and corrections to the model parameters are iteratively fitted to minimize them. This fitting process is repeated with the updated (post-fit) model until convergence is achieved.

In particular, in our analysis we make use of the \texttt{T2} binary model \cite{tempo2manual}, which is a highly flexible and general framework in \texttt{TEMPO2}. Unlike other models, such as \texttt{DDGR}, the \texttt{T2} model does not assume a specific theory of gravity, and instead allows for the inclusion of a wide range of PK parameters and relativistic effects in a more model-independent way. This is convenient for studying alternative theories of gravity since it allow us to set independent constraints on each parameter without enforcing a specific dependence between them, that is, without assuming GR.

In this work, rather than relying only on the standard least-squares approach implemented in \texttt{TEMPO2}, we employ the \texttt{MCMC4Tempo2} \cite{Voisin_2020MCMC} plugin to perform a Bayesian analysis of the timing model parameters. This tool uses a parallelized, affine-invariant Markov Chain Monte Carlo (MCMC) sampler to explore the posterior probability distributions of the parameters given the timing residuals and TOA uncertainties. The result is a statistically robust estimation of the parameters, as well as their full joint probability distributions, which allows for a more complete characterization of uncertainties, correlations, and potential degeneracies.

We have previously seen in Eq.~\eqref{sistemadeeq} how a precise measurement of the orbital period derivative \( \dot{P}_b \) in binary pulsars can be used to constrain both a possible time-variation of the gravitational constant \( \dot{G}/G \) and the strength of dipolar gravitational radiation, parametrized by \( \kappa_D \). In this section, we present an analysis based on a selection of pulsar systems, aiming to provide robust and consistent bounds on these effects.

We analyze these pulsars making use of the most recent data release \footnote{Data sets can be found on the EPTA repository \url{https://gitlab.in2p3.fr/epta/epta-dr2/-/tree/2911d0e52e0c8a4e528c4e3aa46b868ced1910e8/} .} from the European Pulsar Timing Array (EPTA) Collaboration~\cite{EPTAcollab}, which provides TOAs obtained from five radio telescopes across Europe. Each pulsar's TOAs are processed using the software package \texttt{MCMC4Tempo2} discussed previously. For each pulsar, we fit the TOAs using the corresponding ephemeris and run the MCMC analysis, obtaining a chain with approximately 120,000 samples, from which we compute posterior means and standard deviations. The resulting distributions are generally well approximated by Gaussians.

Finally, once we obtain the MCMC chains, then Eq.~\eqref{sistemadeeq} can be solved through the combined analysis of two or more pulsar systems. In our approach, this is done using the previously analyzed pulsars along with the binary system PSR J0437–4715, which serves as an auxiliary reference in the system of equations. This choice is motivated by the fact that the variation in the orbital period due to dipolar emission is stronger in binaries with shorter orbits since $\dot{P}_b^{\text{D}} \, \propto \, P_b^{-1}$, while the variation caused by $\dot{G}$ scales as $\dot{P}_b^{\dot{G}} \, \propto \, P_b$. In order to break the degeneracy between both contributions, it is convenient to solve Eq. \eqref{sistemadeeq} for $\dot{G}/G$ and $\kappa_D$ for the combined analysis of two binary systems with different orbital periods, following the methodology used by \cite{Lazaridis}.

Likewise, we adopt directly the value reported by \cite{VERBIEST04}, who provided a highly precise estimate of
\begin{equation}
\left(\frac{\dot{P}_b}{P_b}\right)^{\text{exc}} = (3.2 \pm 5.7) \times 10^{-19}\ \mathrm{s}^{-1}.
\end{equation}
This choice is motivated by the fact that the observed orbital period derivative in PSR J0437–4715 is heavily dominated by non-intrinsic contributions, particularly kinematic effects such as those described in Eqs. \eqref{pbgalact} and \eqref{shkolvs}. These contributions introduce significant uncertainties when estimated solely from the timing chains, especially in the presence of poorly constrained parallax values which strongly affect the Shklovskii correction. 
This is consistent with the methodology adopted in previous works \cite{Lazaridis, PSRJ1738+0333test}, where PSR J0437--4715 is treated as a reference system due to its well-characterized kinematics and orbital parameters \cite{VERBIEST04, DELLER}. Using well-measured values from the literature for such a system, provides more reliable and conservative constraints on the parameters \(\dot{G}/G\) and \(\kappa_D\).

In this work we also present an end-to-end pipeline, taking into account both \texttt{TEMPO2} and \texttt{MCMC4Tempo2}, but also several publicly available TOAs. Our pipeline, consistently analyzes all the TOAs data and directly produces the results and plots in this paper.

\section{Results}
\label{sec:results}
Based on the combined timing data analysis of PSR J1713+0747 and PSR J0437--4714, we obtain the following $95\%$ confidence level constraints on both the variation of the gravitational constant and the dipolar radiation constant
\begin{align}
\frac{\dot{G}}{G} &= (0.32 \pm 0.31) \times 10^{-12}~\text{yr}^{-1},\label{constraintgdot} \\
\kappa_D &= (-0.04 \pm 0.14) \times 10^{-4}.\label{constraintkd}
\end{align}
These results are in good agreement with those obtained in Ref.~\cite{Zhu_2018}, who reported $
\dot{G}/{G} = (-0.1 \pm 0.9) \times 10^{-12}~\text{yr}^{-1}, \quad \kappa_D = (-0.7 \pm 2.2) \times 10^{-4}$.
While the central value of the varying gravitational constant differs slightly, both results are statistically compatible. Notably, our constraint on the dipolar radiation constant is significantly tighter, by an order of magnitude.

\begin{figure}
\centering
\includegraphics[width=0.308\textwidth]{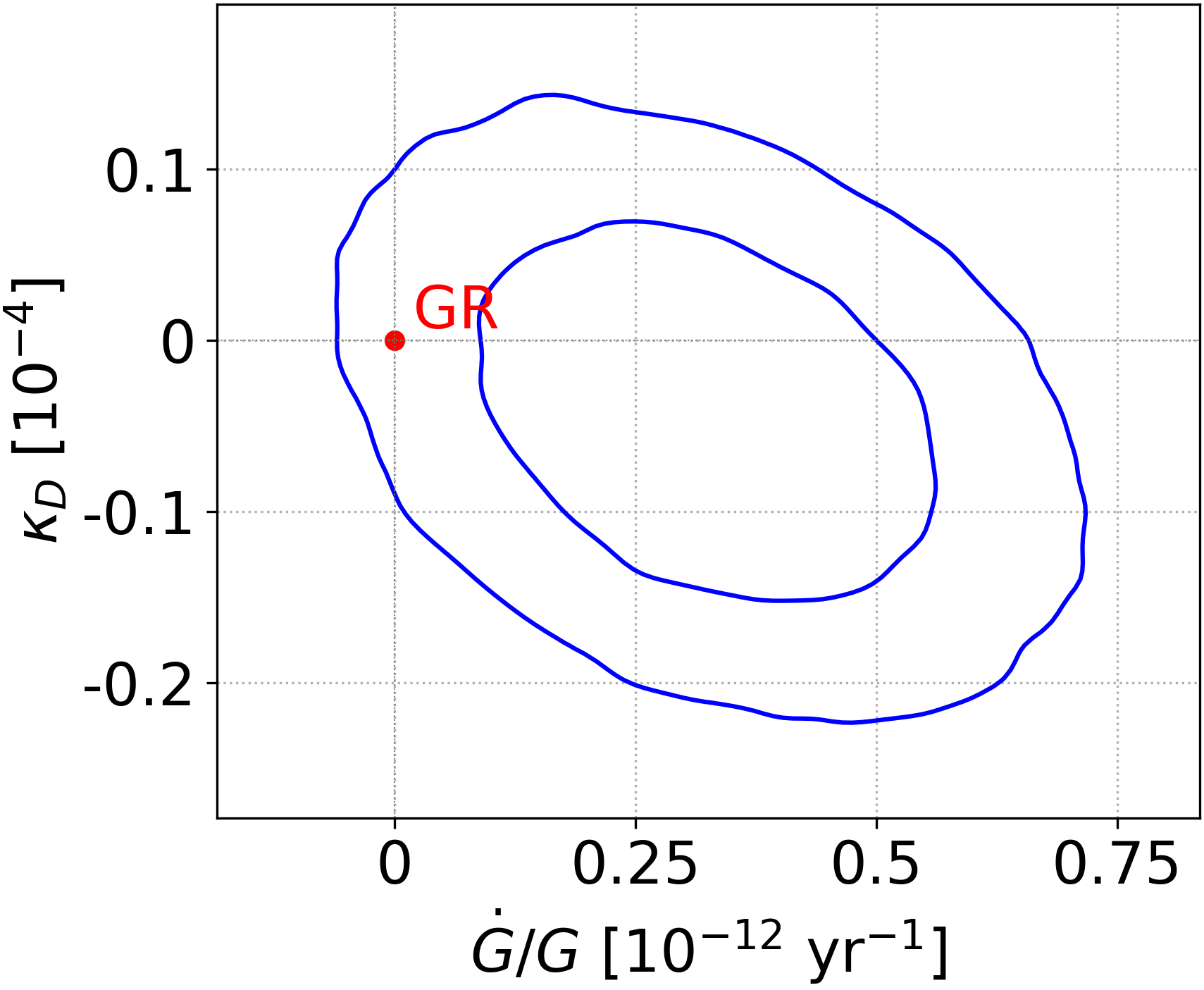}
\includegraphics[width=0.327\textwidth]{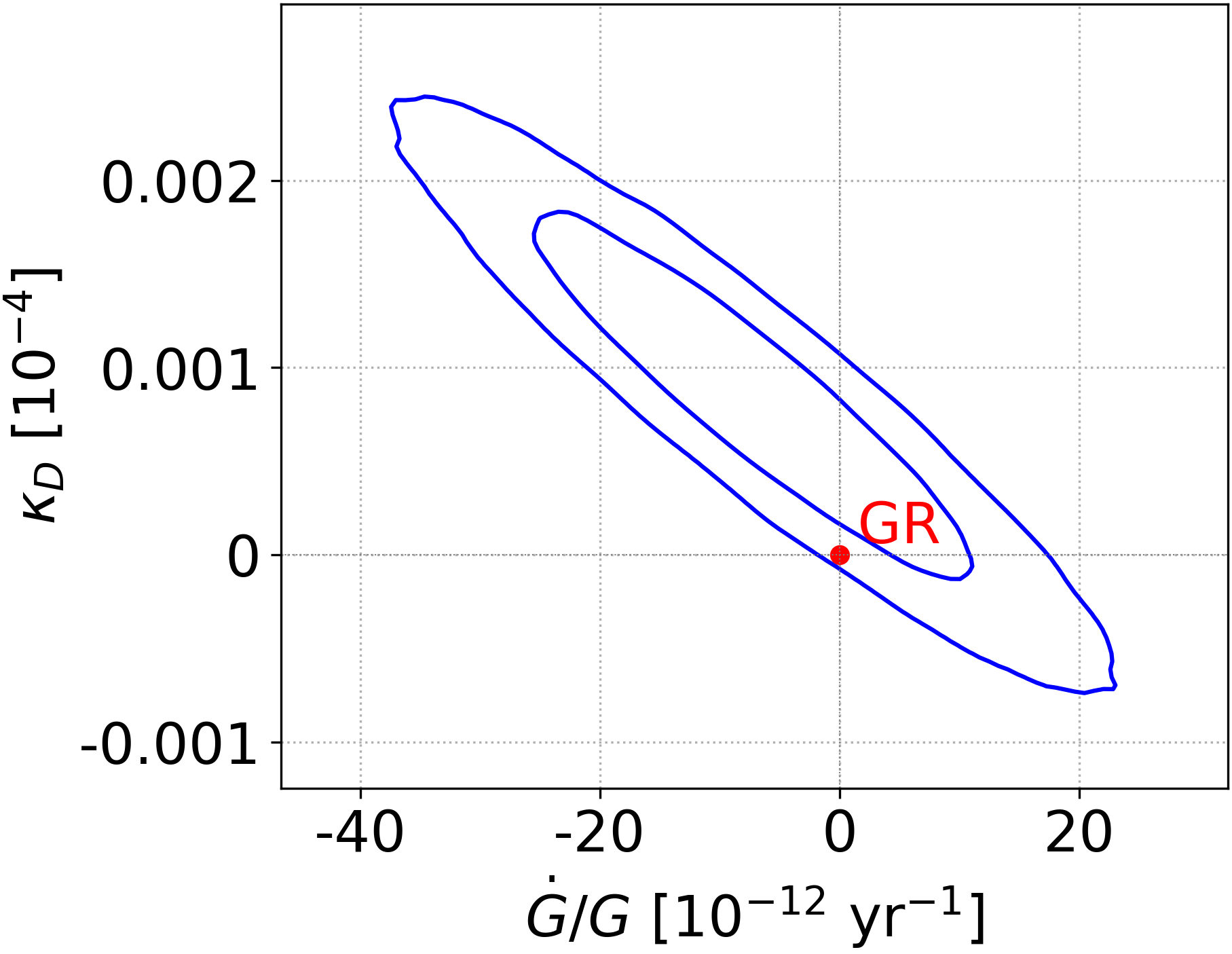}
\includegraphics[width=0.327\textwidth]{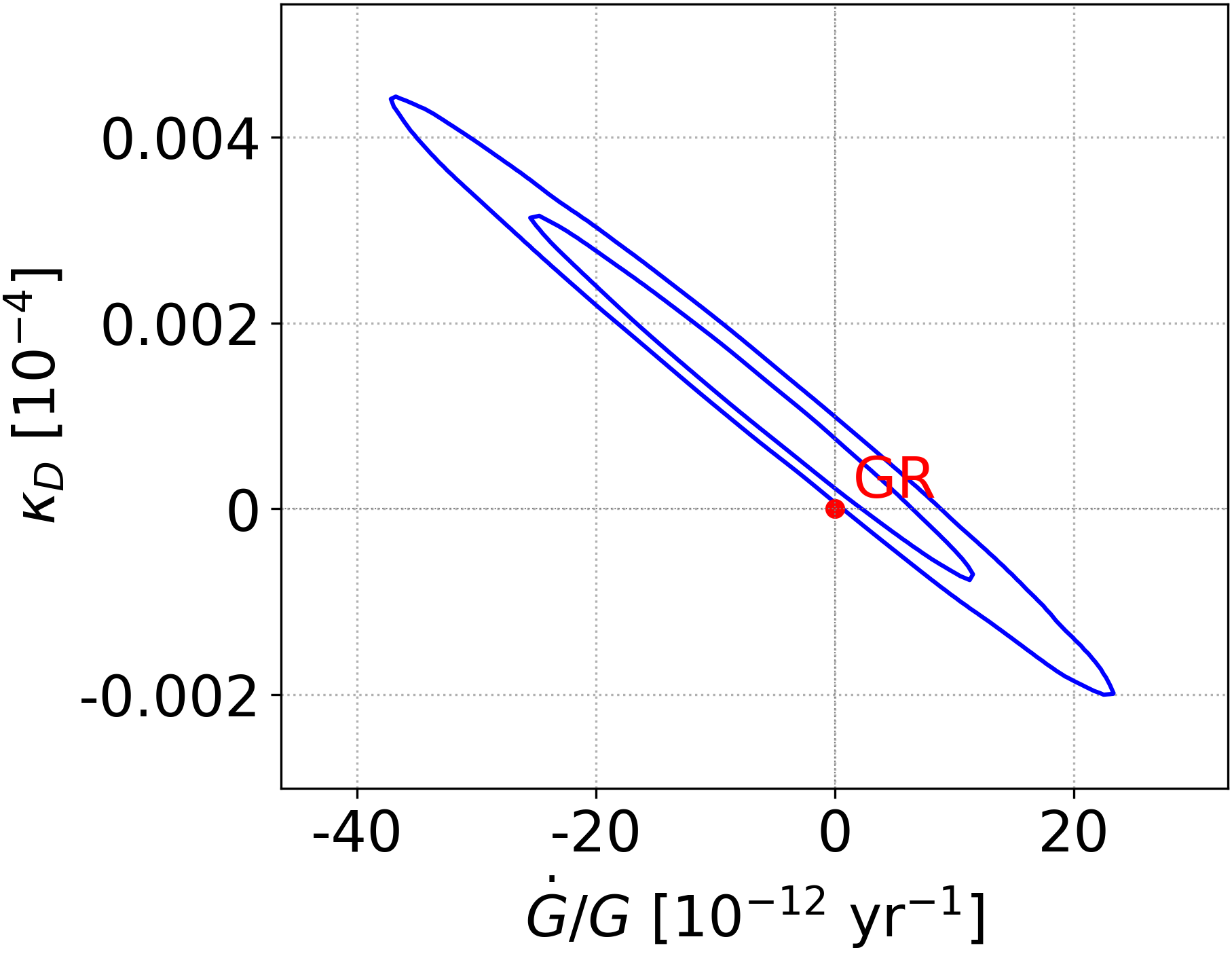}
\caption{Joint constraints on $\dot{G}/G$ and $\kappa_D$ through the combined analysis of the auxiliary system PSR J0437--4715 with the binary systems: PSR J1713+0747 (left panel), PSR J1738+0333 (central panel) and PSR J1012+5307 (right panel). The inner and outer blue contour levels include respectively the $68.3\%$ and $95.4\%$ of all probability. The expected value for GR $(\kappa_D=0, \, \, \dot{G}/G=0)$ is represented by a red dot.}
  \label{J1713GDOT}
\end{figure}

We also present the analysis for the pulsars PSR J1738$+$0333 and PSR J1012$+$5307, obtaining respectively the following $95\%$ confidence level constraints on both the variation of the gravitational constant and the dipolar radiation constant
\begin{equation}
\frac{\dot{G}}{G} = (-7 \pm 24) \times 10^{-12}~\text{yr}^{-1}, \quad \quad \kappa_D = (0.8 \pm 1.3) \times 10^{-7},
\end{equation}
\begin{equation}
\frac{\dot{G}}{G} = (-7 \pm 24) \times 10^{-12}~\text{yr}^{-1}, \quad \quad \kappa_D = (1.2 \pm 2.6) \times 10^{-7}.
\end{equation}
In both cases, we obtain a rather weak constraint on the gravitational constant, about one order of magnitude larger, while the constraint on the dipolar radiation constant is extremely tight, especially when compared to previous bounds obtained for PSR~J1713$+$0747. This could indicate that the degeneracy between $\dot{G}/G$ and $\kappa_D$ has not been fully broken, despite having included an auxiliary pulsar with an orbital period an order of magnitude longer. 

Other significant reason might be related to inaccurate parameter estimates. For the pulsar PSR~J1738$+$0333 we already had to set a fixed parallax obtained by optical measurements, rather than including it as a free parameter on the timing model. Nevertheless, comparing our results to other published works studying these pulsars, which obtain constraints closer to those reported in Eqs.~\eqref{constraintgdot} and \eqref{constraintkd}, we observe a significant discrepancy in the observed orbital period derivative for PSR~J1738$+$0333 with respect to Ref.~\cite{PSRJ1738+0333test}, and a deviation in the parallax measurement of PSR~J1012$+$5307 compared to Ref.~\cite{Lazaridis}, which could affect the results especially through the Shklovskii effect.

\section{Conclusions}
\label{sec:conclusion}
In this work, we have explored the use of binary pulsar systems as precision laboratories for tests for deviations from GR, predicted by alternative theories of gravity, specifically a varying gravitational constant and the emission of dipolar gravitational radiation. The aforementioned effects are related to deviations, from the GR prediction, in the orbital period derivative $\dot{P}_b$ in binary pulsars, via Eqs.~\eqref{Nordvelt} and \eqref{dipolar}. This is done transforming the TOAs of pulsar signals into a common reference frame, by applying a timing model that parametrizes all time delays that might have affected the measured pulse. 

Then, the best fit timing model of the TOAs, gives both the Keplerian and post-Keplerian parameters that describe the binary's orbital dynamics, without assuming any theory of gravity. In particular, we have obtained the observed orbital period derivative $\dot{P}_b^{\text{obs}}$, a key parameter for our analysis. After accounting for different kinematic effects, particularly due to the galactic acceleration and the Shklovskii effect, we obtain its intrinsic value, which is purely relativistic, and that can be compared to the GR prediction via Eq.~\eqref{EinsteinPbdot}.

To do so, we have used the software \texttt{TEMPO2} and its \texttt{MCMC4Tempo2} plugin, along with a \texttt{python} pipeline to unify the codes in a single framework, so as to carry out a Bayesian fit of the timing model parameters, allowing us to estimate the parameters' uncertainties and correlations. We then combined results from selected pulsars with PSR~J0437--4715, used as an auxiliary reference, to obtain constraints on both the time-varying gravitational constant $\dot{G}/G$ and the dipolar emission strength $\kappa_D$.

In particular, the joint analysis of PSR~J1713$+$0747 and PSR~J0437--4715 has provided tight and competitive bounds, setting a more stringent limit on $\kappa_D$ than in previous works. This improvement could be related to the use of recent EPTA data, which has increased the number of TOAs. Moreover, our constraint on $\dot{G}/G$ agrees well with earlier results from binary pulsar timing \cite{Zhu_2018} remaining weaker than the best Solar System bounds, such as those from Lunar Laser Ranging (LLR), which report  $\dot{G}/{G} = (-0.7 \pm 3.8) \times 10^{-13}~\mathrm{yr}^{-1}$ \cite{LLR}. 

However, when extending the analysis to PSR~J1738$+$0333 and PSR~J1012$+$5307, the resulting constraints on $\dot{G}/G$ become significantly weaker while on $\kappa_D$ become highly tight. This is likely due to increased uncertainties in key parameters, such as parallax and $\dot{P}_b^{\text{obs}}$, underlining the sensitivity of the method to precise timing measurements, and potentially requiring a joint analysis incorporating complementary optical observations.
Additional contributions neglected in the calculation of the intrinsic orbital period derivative $\dot{P}_b^{\text{int}}$, such as tidal or mass loss effects from the pulsar or its companion, may contribute more significantly than initially expected, potentially biasing the derived constraints. Ultimately, the obtained constraints might indicate that the degeneracy between both parameters has not been fully broken.

Related to the current published data, we have noted that Ref.~\cite{Lazaridis} initially sets $\dot{P}_b^D = 0$ for PSR~J0437--4715 in the combined analysis of Eq.~\eqref{sistemadeeq}. This allows them to isolate the contribution from a varying gravitational constant $\dot{G}/G$ more cleanly. In our main analysis, by contrast, we did not impose any such assumption, treating both $\dot{G}/G$ and $\kappa_D$ as free parameters. While this choice may introduce greater degeneracy between the parameters, it avoids possible biases since any unmodeled dipolar contribution would be effectively absorbed in $\dot{G}/G$.
Other works in the literature \cite{PSRJ1738+0333test, Zhu_2018} that refer to the methodology of Ref.~\cite{Lazaridis} when including PSR~J0437--4715 in their joint analyses do not explicitly state whether they set the dipolar radiation term to zero in their treatment. To explore the effect of this modeling choice, we also repeated our analysis following the same prescription. Under this assumption, our constraints for PSR~J1713$+$0747 on both $\dot{G}/G$ and $\kappa_D$ become slightly weaker, while the GR prediction lies well within the $1\sigma$ region. For PSR~J1012$+$5307 and PSR~J1738$+$0333 we obtained the same results as before.

Finally, we remark that a constraint on the time-varying gravitational constant can be interpreted within the framework of modified gravity theories. In particular, due to the rescaling of the effective gravitational constant by the scalar field $\phi$ added in scalar-tensor theories Eq. \eqref{gdotSCALARTENSOR} and naturally arising from compactified extra dimensions Eq. \eqref{GDOTEXTRADIM}, this limit can be translated into a bound for $\dot{\phi}/\phi$. That is, observationally obtained constraints can be directly translated into limits on the dynamics of the underlying fields in these theoretical scenarios.
Our $\kappa_D$ test is a generic test of dipolar radiation, included for the purpose of generalizing the constraint on $\dot{G}/G$ for general SEP-violation theories. For a more detailed discussion on this effect, one should take into account the nature of the theory on study  \cite{PSRJ1738+0333test}, since the $\mathcal{O}(s_p^3)$ terms that we have neglected in Eq. \eqref{dipolar} may be significant even at leading order for some cases.
Even so, from the obtained results it is worth noting that, although dipolar radiation arises at 1.5PN order, it is strongly suppressed. Together with the limits of the time-varying gravitational constant, this  reinforces the validity of GR within the current observational precision.

Overall, these results highlight the power of binary pulsar timing in testing gravitational theories.  While some of the derived bounds are limited by astrophysical, observational uncertainties and model degeneracies, particularly for certain pulsars, future improvements in parallax measurements and timing precision may lead to significantly better constraints. Continued monitoring of known systems, as well as the discovery and characterization of new binary pulsars, will play a crucial role in refining these tests of gravity.

\acknowledgments
The authors acknowledge the use of the Hydra cluster at the IFT, and the use of the \texttt{TEMPO2} and \texttt{MCMC4Tempo2} codes. SN and MC also acknowledge support from the research project PID2021-123012NB-C43 and the Spanish Research Agency (Agencia Estatal de Investigaci\'on) through the Grant IFT Centro de Excelencia Severo Ochoa No CEX2020-001007-S, funded by MCIN/AEI/10.13039/501100011033. MC acknowledges support from the ``Ram\'on Areces'' Foundation through the ``Programa de Ayudas Fundaci\'on Ram\'on Areces para la realizaci\'on de Tesis Doctorales en Ciencias de la Vida y de la Materia 2023'' and the hospitality of the Max Planck Institute for Gravitational Physics in Potsdam, during the period in which part of this work was completed.

\paragraph{Code availability:}
The \texttt{python} code for the numerical pipeline used in our analysis of the TOAs and the production of all the figures in the paper, will be made publicly available at \url{https://github.com/sofiabussieres/Constraints-on-modified-gravity.git}, upon publication. 

\bibliography{biblio.bib}

\end{document}